\def\etal{et al.\ }

\def\spose#1{\hbox to 0pt{#1\hss}}
\def\approxlt{\mathrel{\spose{\lower 3pt\hbox{$\sim$}}
	\raise 2.0pt\hbox{$<$}}}
\def\approxgt{\mathrel{\spose{\lower 3pt\hbox{$\sim$}}
	\raise 2.0pt\hbox{$>$}}}
	
\def\Mdot{\hbox{$\dot M$}}
\def\degmark{$^\circ$}
\def\<{\thinspace}
\def\s{\hbox{\phantom{5}}}	

\def\cm{{\rm\thinspace cm}}
\def\ct{{\rm\thinspace ct}}
\def\erg{{\rm\thinspace erg}}

\def\K{{\rm\thinspace K}}
\def\keV{{\rm\thinspace keV}}

\def\km{{\rm\thinspace km}}
\def\kpc{{\rm\thinspace kpc}}

\def\Mpc{{\rm\thinspace Mpc}}
\def\Msun{\hbox{$\rm\thinspace M_{\odot}$}}

\def\s{{\rm\thinspace s}}
\def\yr{{\rm\thinspace yr}}

\def\ctps{\hbox{$\ct\s^{-1}\,$}}

\def\pcmcu{\hbox{$\cm^{-3}\,$}}

\def\ergps{\hbox{$\erg\s^{-1}\,$}}

\def\kmps{\hbox{$\km\s^{-1}\,$}}

\def\Msunpyr{\hbox{$\Msun\yr^{-1}\,$}}

\def\pcmsq{\hbox{$\cm^{-2}\,$}}

\def\pcmcuK{\hbox{$\cm^{-3}\K$}}

\def\kmpspMpc{\hbox{$\kmps\Mpc^{-1}$}}

\documentstyle[psfig]{mn}  
\begin{document}

\title{Detection of X-ray emission from the host clusters of 3CR quasars}

\author[C.S. Crawford et al ]
{\parbox[]{6.in} {C.S. Crawford$^{1}$, I. Lehmann$^2$, A.C. Fabian$^1$,
M.N. Bremer$^3$ and G. Hasinger$^2$ %
\\
\footnotesize
1. Institute of Astronomy, Madingley Road, Cambridge CB3 0HA \\
2. Astrophysikalisches Institut Potsdam,  An der Sternwarte 16, D-14482 Potsdam, Germany\\
3. Dept. of Physics, University of Bristol, Tyndall Avenue, Bristol BS8 1TL \\}}

\maketitle
\begin{abstract}
We report the detection of extended X-ray emission around several powerful
3CR quasars with redshifts out to 0.73. The ROSAT HRI images of the
quasars have been corrected for spacecraft wobble and compared with an
empirical point-spread-function. All the quasars examined show excess emission at
radii of 15 arcsec and more; the evidence  being strong for the more
distant objects and weak only for the
two nearest ones, which are known from other wavelengths not to lie
in strongly clustered environments. The spatial profiles of the extended
component is consistent with thermal emission from the intracluster
medium of moderately rich host clusters to the quasars. The total
luminosities of the clusters are in the range $\sim 4\times 10^{44}-
3\times 10^{45}\ergps$, assuming a temperature of 4~keV. The inner
regions of the intracluster medium are, in all cases, dense enough to
be part of a cooling flow.
\end{abstract}


\section{Introduction}

The two most powerful FR~II radio sources in the nearby Universe -- Cyg~A
and 3C295 -- are each located at the centre of a dense, moderately rich
cluster of galaxies.  While such an environment is exceptional for a
low-redshift FR~II galaxy, it appears to be common around powerful radio
objects at earlier epochs. Above a redshift of 0.5, radio-loud objects (both
the quasars and radio galaxies) are inferred to lie in clusters of galaxies
of moderate optical richness. The evidence for such an environment includes
optical and near-IR galaxy counts (Yee \& Green 1987; Yates et al 1989; Hill
\& Lilly 1991; Ellingson et al 1991; Dickinson 1997), high gas pressures
within a radius of 30\kpc\  (Crawford \& Fabian 1989; Forbes et al 1990; Bremer
et al 1992; Durret et al 1994), cD-type host galaxy profiles (Best et al
1998), a gravitational arc (Deltorn et al 1997), and 
 lensing shear of surrounding field galaxies (Bower \& Smail
1997). The properties of the radio source itself also imply the presence of
a confining medium: a large-scale working surface on which the jets form the
radio lobes; a steep radio spectrum; and a high minimum pressure in regions
of relaxed radio structure (Bremer et al 1992). A Faraday depolarization
asymmetry (Garrington \& Conway 1991), the distortion and compression of
high-redshift radio source morphologies (Hintzen et al 1983; Barthel \& Miley
1988) and sources with very high Faraday rotation measures (Carilli et al
1994; Carilli et al 1997) all corroborate the inference of a dense, clumpy
medium surrounding the radio source. Thus it appears that the deepest
potential wells we can readily pinpoint at $z\ge1$ are those around powerful
radio sources.

The cluster distribution at high redshift can provide a stringent
cosmological test (see e.g. Donahue et al 1998), and can also be
compared to the X-ray luminosity function of clusters at low redshift
(eg Ebeling et al 1997). Whilst it may result in a sample of clusters
biased to only those that can host an active nucleus, using radio
sources to identify the location of deep potential wells is a
promising way of finding clusters out to and beyond a redshift
$z\sim1$ (Crawford 1997). Current X-ray surveys of clusters detected
from the ROSAT All-Sky Survey (eg Ebeling et al 1998) do not reach
sufficiently faint flux levels, and studies of deep serendipitous
X-ray pointings (e.g. Rosati et al 1998) cover only a small fraction
of the sky. The first step, however, is simply to confirm that
powerful radio quasars beyond a redshift of a half really do lie at
the centre of clusters of galaxies.

The clearest way to determine directly the presence of a cluster of
galaxies is to detect thermal X-ray emission from its hot intracluster
medium. A certain degree of success has been achieved in detecting and
spatially resolving the X-ray emission around distant ($0.5<z<2$)
radio galaxies using ROSAT (Crawford \& Fabian 1993; Worrall \etal
1994; Crawford \& Fabian 1995a, 1996a,b;
Crawford 1997; Dickinson 1997; Hardcastle, Lawrence \& Worrall 1998;
Carilli et al 1998). Any X-rays emitted by the central bright nucleus
of radio galaxies are assumed to be absorbed along the line of sight,
as observed for the powerful low-redshift radio galaxy Cygnus-A (Ueno
et al 1994). The inferred bolometric luminosity of the X-ray sources
associated with the radio galaxies is $\sim0.7-18\times10^{44}$\ergps,
easily compatible with that expected from moderately rich clusters of
galaxies around the radio sources. There could also be a contribution
to the extended X-ray emission from inverse Compton scattering of the
hidden quasar radiation (eg Brunetti, Setti \& Comastri 1997).

In the case of radio quasars, however, the X-ray detection of the spatially
extended environment is complicated by the presence of bright
spatially-unresolved X-ray emission from the active nucleus.  The ROSAT PSPC
did not combine the necessary sensitivity with a sufficiently good
point-response function, needed to both detect and resolve any cluster
emission around quasars. Upper limits of $1.6-3.5\times10^{44}$\ergps (in
the 0.1-2.4\keV\ rest-frame band) to X-ray emission from the environment of
three radio-loud quasars have been derived from ROSAT HRI data (Hall et al
1995, 1997) assuming the cluster emission profile is modelled by a King law.

We have also obtained ROSAT HRI data to spatially resolve and detect
the extended emission from the intracluster medium around each of a
small sample of intermediate-redshift radio-loud quasars. The
detection of such a component is, however, complicated by the wobble
of the spacecraft during the observation. This occurs on a $\sim402$~s
period, and when the attitude of the spacecraft is not well
reconstructed, leads to smearing of the point-spread function (PSF). The bright emission from
the quasar nucleus can then contaminate the outer regions where we
hope to detect emission from any surrounding cluster, and this has so
far hindered our progress in interpreting the data.  In this paper, however, 
we present an analysis of our ROSAT HRI data taken of
seven intermediate-redshift (0.1$<z<$0.8) radio-loud quasars, which employs a
new correction for the spacecraft wobble derived by Harris et al.
(1998).

A contemporaneous and independent analysis of an overlapping dataset
using this technique has been carried out by Hardcastle \&
Worrall (1999), who obtain similar results.

\section{Observations and analysis}

We use the ROSAT data of intermediate-redshift, radio-loud quasars for
which there is prior evidence from other wavebands for a cluster
environment (see notes on individual quasars for details). We
preferentially selected quasars of only moderate X-ray luminosity in
order to minimise the contrast between the nuclear emission and any
cluster emission. These targets were supplemented by data available
from the ROSAT public archive on 3C273 and 3C215. We also include the
observations of H1821+643 to form a comparison to the results of Hall
etal (1997). The observations used, and details of the quasars are
listed in Table~\ref{tab:obslog}.

\onecolumn
\begin{table}
\caption{Target sample and results \label{tab:obslog}}
\begin{tabular}{lccccccc}
     & &  &  &  &  &   & \\
     & &  &  &  &  &   & \\
Quasar& RA         & DEC      & Redshift & N$_H$ & ROR    & Exposure &  Roll angle \\
      & (J2000)    & (J2000)  & $z$      & ($10^{20}$\pcmsq) & & (sec)  &  interval   \\

3C48 & 01 37 41.3  & 33 09 35  & 0.367 & 4.54 & 800634n00 & 37362 & 1-60000 \\
3C215 & 09 06 31.9 & 16 46 13 & 0.412  & 3.65 & 800753a01 & 39173 & 27000-58000 \\
      &            &          &        &      & 800753n00 & 17231 & 14000-22000 \\
      &            &          &        &      & 800718n00 & 16148 & 1-50000       \\
3C254 & 11 14 38.5 & 40 37 20 & 0.734  & 1.90 & 800721n00 & 29162 & 29000-49000 \\
3C273 & 12 29 06.7 & 02 03 09 & 0.158  & 1.79  & 701576n00 & 68154 & 1-32000  \\
3C275.1& 12 43 57.7 & 16 22 53 & 0.555 & 1.99 & 800719n00 & 25396 & 9000-32000 \\
3C281 & 13 07 53.9 & 06 42 13 & 0.602  & 2.21 & 800635a01 & 20299 & 12000-24000 \\
      &            &          &        &      & 800635n00 & 18220 & 1-14000 \\ 
3C334 & 16 20 21.5 & 17 36 29 & 0.555  & 4.24 & 800720n00 & 28183 & 17000-42000 \\
H1821+643 & 18 21 57.3 & 64 20 36  & 0.297 & 4.04 & 800754n00  & 29427 & 23000-31000 \\

     & &  &  &  &  &   & \\
\end{tabular}
\\
Notes:\\
Further possible observations of 3C215 (800718a01), 3C194 (800803n00)  and 3C280 (800802n00) all had too few
photons in the source for a satisfactory  wobble correction. \\
The hydrogen column density along the line of sight to
each quasar  (N$_H$) is calculated from the data of Stark et al (1992). \\
ROR in column 6 is the ROSAT observation request sequence number. \\
The roll angle interval tabulated in the final column is given in sequence
numbers taken from the attitude file, which contains the attitude
information of the telescope (such that roll angle x1 at time y1 gives
sequence number 1). The time interval of the spacecraft clock is 1\s.
Sequence numbers from 1 to 34000 thus mean that the data is extracted from
between the time y1 and the time y34000 (34000\s). \\

\end{table}
\twocolumn

\subsection{Wobble-correction technique}

The spatial analysis of ROSAT HRI observations is often complicated by 
smearing of the image on the order of 10 arcsec (Morse 1994). This
degradation of the instrinsic resolution of the HRI instrument (5
arcsec) can be induced by errors in the aspect solution associated
with the wobble of the ROSAT spacecraft, or with the reacquisition of
the guide stars. To counteract this effect, we use the 
wobble-correction technique of Harris et al (1998) which minimizes the spatial smearing of the
sources.

 The technique is based on the simple assumption that in the case of a
 stable roll angle (i.e. the same guide star configuration) the aspect
 error is repeated through each cycle of the wobble. We thus select
 data only from the longest constant roll angle interval of an 
 observation and folded these data over the ROSAT 402~s wobble
 phase. This phase was grouped into a number of intervals (5, 10 or
 20) in order to calculate phase-resolved sub-images. The centroid of
 each of the sub-images was calculated by fitting a 2d-gaussian to the
 brightest source near the field center. The wobble-corrected HRI
 events list is reconstructed by adding the sub-images which have been
 shifted to the centroid position of the uncorrected HRI image. If a
 sub-image contains too few photons with which to determine the
 centroid, then the position of the sub-image is not changed. This
 makes sure that source extension cannot be produced by a failure of
 the centroid determination in one or more sub-images.

For X-ray sources with countrates $\sim$0.1\ctps the method can
reduce the full width half maximum by about 30 per cent (cf. Harris et al.
1998). The efficacy of this technique has been tested on HRI observations of
low-luminosity X-ray AGN, some of which show no sign of extended X-ray
emission once corrected for wobble (Lehmann et al. 1999). We cannot apply
the wobble-correction technique to all possible data on suitable targets in
the literature, as it can only be applied to observations that have
a sufficient number of photons within a constant spacecraft roll angle (cf.
Table~\ref{tab:obslog}).

\subsection{Radial profile analysis}

Our aim is to search for spatially extended X-ray emission originating
in any intracluster medium around the quasars.
We have derived a background-subtracted radial profile from the un-corrected
and wobble-corrected HRI data of the X-ray source associated with each quasar.
First we determined the centroid position of the X-ray source by fitting a
2D Gaussian. Then we calculated the counts per arcsec$^2$ within the rings around
the centroid
position from 0 arcsec to 2.5 arcsec in steps of 0.5 arcsec, from 
2.5 arcsec to 15 arcsec in steps of 2.5 arcsec, from 15 arcsec to 100 arcsec
in steps of
12.5 arcsec and from 100 arcsec to 1000 arcsec in steps of 100 arcsec. The
background value was calculated as the median from the 6 rings between 
300 arcsec and 900 arcsec. Finally we subtracted the background value from 
each ring. 

\begin{figure}
\psfig{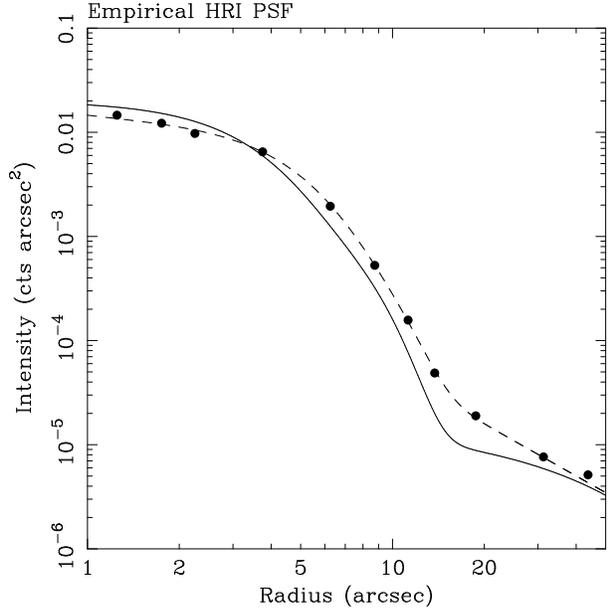}
\caption{ \label{fig:emppsf}
The best-fit analytic model (dashed line) to the empirically-derived HRI PSF (circle
markers), shown with an arbitrary flux scaling. The solid line shows
the standard HRI PSF of David et al (1995) for comparison. }
\end{figure}

First, however, we need to accurately model the HRI PSF which
will allow us to remove the contaminating spill-over of light from the
bright quasar nucleus. The standard HRI PSF of David et al (1995) does not
provide a good fit to the radial profile of observed point sources
(see Fig~\ref{fig:emppsf}). 
This deviation is particularly acute where the PSF shows a sharp drop
at radii between 10 and 30 arcsec, where we expect the contrast
between the nuclear source and any extended cluster emission to begin
to show. Instead we have determined a good analytical characterization
to an empirical PSF derived from observations of 21 ROSAT Bright
survey stars, each of which has undergone the same wobble correction
procedure as the quasar data.  Our best fit is the sum of two
Gaussians and a power-law component (Fig~\ref{fig:emppsf}). Assuming
that this PSF forms a good model for the spillover of quasar nuclear
light, we fix the relative normalizations and widths of the three
components and leave only the overall normalization $n$ of this profile as
the only free variable in the function: 
$$ I(r) = n \{ e^{-0.5(r/4.5001)^2} +  4.376 e^{-0.5(r/2.8644)^2} + 0.9346 r^{-1.6569} \}$$


We follow the same procedure for the profile fitting of each quasar.
First we fit the profile by the empirically-derived HRI PSF alone,
allowing its normalization to vary freely. We then fit this PSF (with
normalization still free to vary) in combination with each of two
models chosen to represent any extended emission. The first model is a
broken power-law, with slope $r^{-1}$ for radii $r<R$ and $r^{-2.1}$
for $r\ge R$, where the break $R$ and the absolute normalization of
the extended component are free to vary. This is an approximation to
the X-ray surface brightness profile of the gas in a typical cluster
of galaxies containing a cooling flow, and $R$ then corresponds to the
cooling radius (eg Crawford \& Fabian 1995b). The second model employed is a projected King law,
with index fixed at -1.5, and the core radius $R$ and normalization
left as free parameters. Given the errors inherent in whether such
simple models truly characterize the extended emission, we do not
convolve the extended emission models with the PSF. The relative
normalization between the PSF and extended components are not always
very well determined, so we also derive what should be regarded as a
lower limit to the presence of any extended component by assuming the
nuclear emission accounts for all the light in the X-ray core. We fit
the PSF to the quasar radial profile within the inner 1--5 arcsec and
then subtract this model and fit the residuals by each of the cluster
models. We execute these 5 model fits to the profiles out to a radius
of 50 arcsec (11 data points), yielding 10, 8, and 9 degrees of
freedom for the psf only, psf+extended component models, and the fit
of the extended component model to the residual after subtraction of
the normalised HRI PSF. We then repeat the fits to the profile out to
a radius of 100 arcsec (15 data points), yielding 14, 12 and 13
degrees of freedom to the fits as above. The fitting analysis is
carried out first for each quasar image in the absence of any wobble
correction, and then for the images corrected using different phase
intervals.

\begin{figure*}
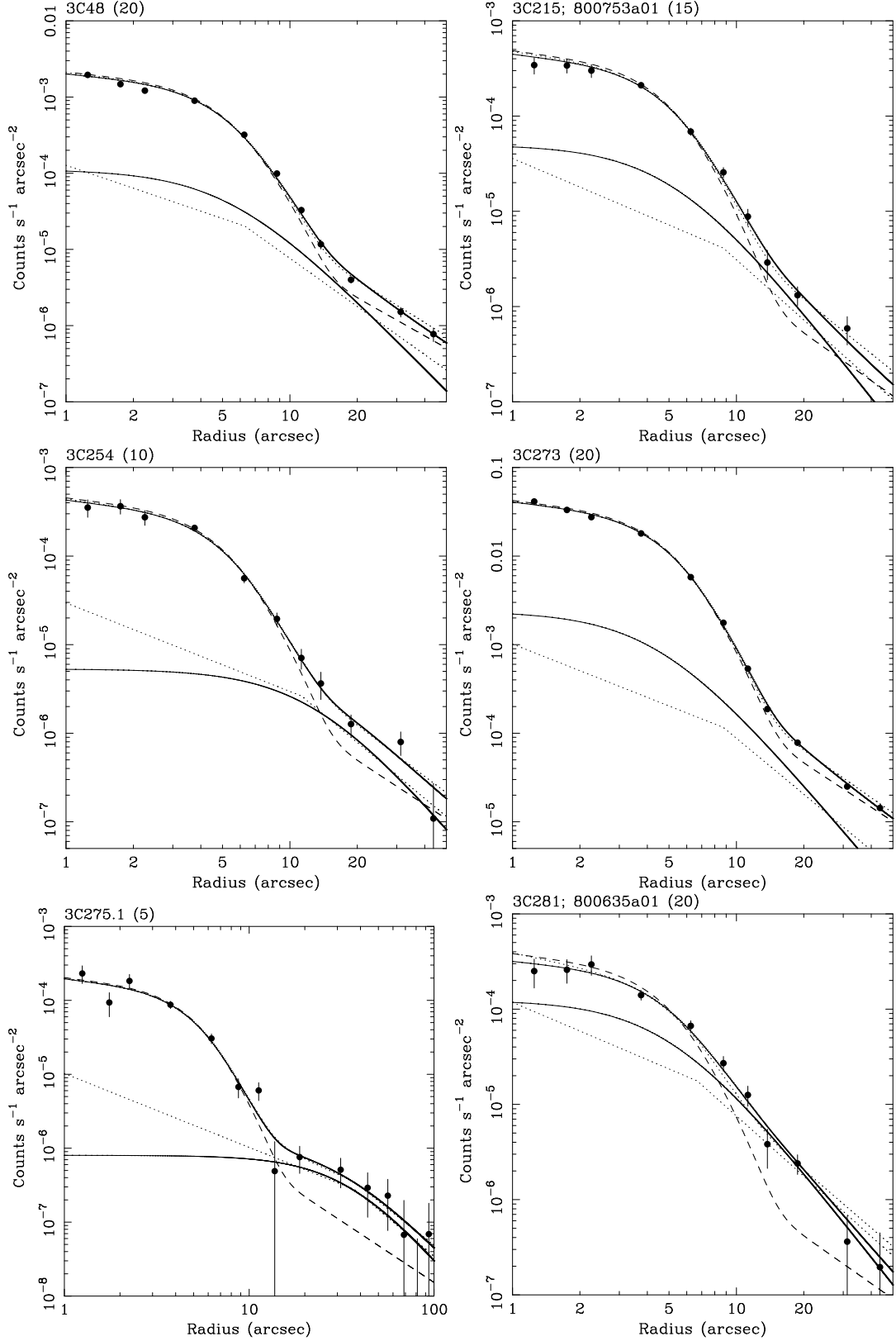

\vbox{
\hbox{
\psfig{figure=3c48prof.ps,width=0.4\textwidth,angle=270}
\psfig{figure=3c215prof.ps,width=0.4\textwidth,angle=270}
}\hbox{
\psfig{figure=3c254prof.ps,width=0.4\textwidth,angle=270}
\psfig{figure=3c273prof.ps,width=0.4\textwidth,angle=270}
}\hbox{
\psfig{figure=3c275.1prof.ps,width=0.4\textwidth,angle=270}
\psfig{figure=3c281prof.ps,width=0.4\textwidth,angle=270}
}}
\caption{ \label{fig:profs}
The radial profile of the X-ray emission from each of the quasars out
to 50 arcsec, with the exception of 3C275.1 where the profile
is shown out to 100 arcsec.  The phase interval used for this radial
profile is shown in brackets after the quasar name in
each plot. 
The data are shown as solid
circle markers, with the  best fit of the empirical HRI PSF on its own plotted as
a dashed line. 
The best-fit models of the PSF+King law and the PSF+broken power-law
model (upper solid and dotted lines respectively) are plotted, as well
as just the 
extended component to each of these fits (the lower solid and dotted
lines respectively). The fits shown in these plots are marked by bold
font in Table~2.}
\end{figure*}

The detailed results are summarized in Table~2, where the
reduced-$\chi^2$ is given for each fit. We tabulate the best-fit
parameters of the profile fits out to a radius of 50 arcsec and then 100
arcsec in turn: $R$ (in arcsec) representing either the break in the
broken power-law model, or the core radius in the King law model; the
integrated luminosity from the extended component as a percentage of
the total luminosity of the X-ray source; the X-ray luminosities (in
the observed 0.1-2\keV\ energy band) of the quasar component
($L_X^{QSO}$) and that of the cluster component ($L_X^{cl}$) assuming
a power-law of photon index 2 and thermal bremsstrahlung emission at a
temperature of $kT=4\keV$ respectively. (At the redshift of our quasars
this observed band carries about half of the bolometric luminosity for
the thermal spectrum.) The errors are derived from propagating the
$\Delta\chi^2=1$ confidence limits of the fit parameters. Errors are
not shown when the fit was insufficiently robust to extract errors on
all parameters of interest. Table~2, however, demonstrates the full
range of values obtained from the ten model fits employed for each of
the phase intervals and allows one to assess the variation of each
parameter from the systematic uncertainties of PSF normalization and
extended component model employed. A comparison of some of the better
fits to the radial profile of each quasar (those shown in bold font in
Table~2) are displayed in Figure~\ref{fig:profs}. These plots clearly
show that there are significant differences between the PSF-only fit
to the profile, and the fits that include a model for extended
emission. In all this analysis we necessarily assume that any extended
component is both centred on the quasar (in no case do we see any
evidence for a secondary off-centre peak), and derive its properties
such as scale and luminosity assuming that it is at the redshift of
the quasar.

The present data cannot rule out a contribution to the extended
component of X-ray emission from the active nuclei of close companion
galaxies to each of the quasars. Such emission would, of course,
provide further support for a clustered environment. 
The probability of getting an unassociated X-ray source within
an aperture of 1 square arcminute centred on a quasar is less than
$10^{-3}$, at the flux level of the extended emission. Thus there is
little chance of the extended emission component being due to
contamination by fore- or back-ground sources.

\section{Results for individual quasars} 

\subsection{3C48}
Given its proximity to a very luminous source of photoionizaton, the
low ionization state observed in the spatially extended oxygen line
emission around this 3C48 led Fabian et al (1987) to deduce a high
density environment around this quasar. The inferred gas pressure of
3-8$\times10^5$\pcmcuK\ within 30\kpc\ of the quasar core is consistent
with confinement of the extended emission-line region by an
intracluster medium There is, however, no strong evidence for a rich
cluster of galaxies from optical images (Yee, Green \& Stockman 1986; Yates
etal 1989).


The fit to the radial profile is substantially improved by the
addition of an extended component, the best fits being obtained in all
cases when this is represented by a King law. 
 The extended component requires a very consistent value
for the core radius $R$ of around 5-6 arcsec in all fits (1 arcsec
corresponds to 6.2\kpc\  at the redshift of the quasar\footnote{We
assume a value for the Hubble constant of
$H_0=50$\kmpspMpc\ and a cosmological deceleration parameter of
$q_0=0.5$ throughout this paper.}), and accounts
for 10-16 per cent of the total X-ray source. The full variation of
its luminosity is $5-10\times10^{44}$\ergps, with most of the
values derived being to the lower end of this range.

\subsection{3C215}
This quasar lies in a densely clustered environment (Ellingson \etal
1991; Hintzen 1984), and the radio source has a very complex structure
suggestive of deflection and distortion of the radio jet to the
south-east by some external medium. The two sides of the radio source
show asymmetric Faraday depolarization, which can be interpreted as
due to differing lines of sight through a depolarizing cluster medium
(Garrington, Conway \& Leahy 1991). Crawford \& Fabian (1989) inferred
a high gas pressure of over $3\times10^{5}$\pcmcuK\ from the ionization
state of the extended line emission within 30\kpc\ of the quasar
nucleus.

We have extracted radial profiles from three observations of 3C215.
Smearing of the image seems to have badly affected observation number
800753n00, as $R$ decreases substantially after correction of the
image. In all cases, the reduced $\chi^2$ of the fit improves from the
addition of an extended component to the HRI PSF model, although there
is little preference shown between the King and broken power-law
models.  In the fits allowing free
normalization of the PSF component the $R$ derived is in the range 4-9
arcsec (where 1 arcsec corresponds to 6.5\kpc\  at the redshift of the
quasar), with the broken power-law model always yielding the larger
values of $R$. The luminosity of the extended component ranges over
1.6-6.3$\times10^{44}$\ergps, and account for 11-40 per cent of the
total X-ray emission from this source; the higher values are derived
from the King law models.

\subsection{3C254}
3C254 is a quasar with a very asymmetric radio source, and was
discovered by us (Forbes et al 1990) to lie in a spectacular
emission-line region extending out to radii of 80\kpc. The optical
nebula is again at low ionization, with an 
inferred pressure of $>10^6$\pcmcuK\ 
within 30kpc (Forbes et al 1990;
Crawford \& Vanderriest 1997). The kinematics and distribution of the
line-emitting gas in combination with the radio morphology strongly
suggest that the radio plasma to the east of the quasar is interacting
with a dense clumpy environment (Bremer 1997; Crawford \& Vanderriest
1997).  The radio source itself also shows asymmetric depolarization 
(Liu \& Pooley 1991).
 Optical continuum images show an overdensity of faint objects
around the quasar consistent with a location in a compact cluster or
group (Bremer 1997). The HRI image of 3C254 shows a detached secondary
source of X-ray emission approximately 35 arcsec east of the quasar,
but at a very low fraction of the total quasar luminosity. There is no
optical counterpart to this source on the Space Telescope Science
Institute Digitized Sky Survey.

The fitting of the profile is improved by the addition of an extended
component to the HRI PSF model, although it cannot discriminate
between a King or broken power-law model. The extended emission
contains 12-19 per cent of the total X-ray luminosity of the source.
The characteristic scale length $R$ of the extended emission component
varies over 9-15 arcsec (1 arcsec corresponds to 8.1\kpc\ at the
redshift of the quasar), with a luminosity of
5-9$\times10^{44}$\ergps.

\subsection{3C273}
3C273 is famous for its jet, which can be seen in the outermost
contours of our corrected images. The jet is known to emit X-rays at
around 16 arcsec (nearer the core than the radio and optical features
of the jet), but the emission from the jet contributes less than 0.5
per cent of the total X-ray luminosity of the source associated with
the quasar (Harris \& Stern 1987). From optical images, 3C273 may be a
member of a poor cluster of galaxies (Stockton 1980).


All the wobble-corrected profiles of 3C273 require the addition of an
extended component (preferably a King law). Each of the models for the
extended emission yields slightly (and consistently) different
results: the broken power-law fits tend to have $R$ of 8.7 arcsec
(where 1 arcsec corresponds to 3.6\kpc\ at the redshift of the quasar)
and contain 5 per cent of the total X-ray luminosity, at
7$\times10^{44}$\ergps. The King law fits show a greater variation on
the parameters, but have $R$ of only $\sim$4 arcsec, and $\sim11$ per
cent of the total luminosity at 1.3-2.0$\times10^{45}$\ergps. The
extended emission we find in the environs of this quasar is
sufficiently luminous that it cannot easily be ascribed to the jet.

\twocolumn 

\addtocounter{figure}{-1}
\begin{figure}
\psfig{figure=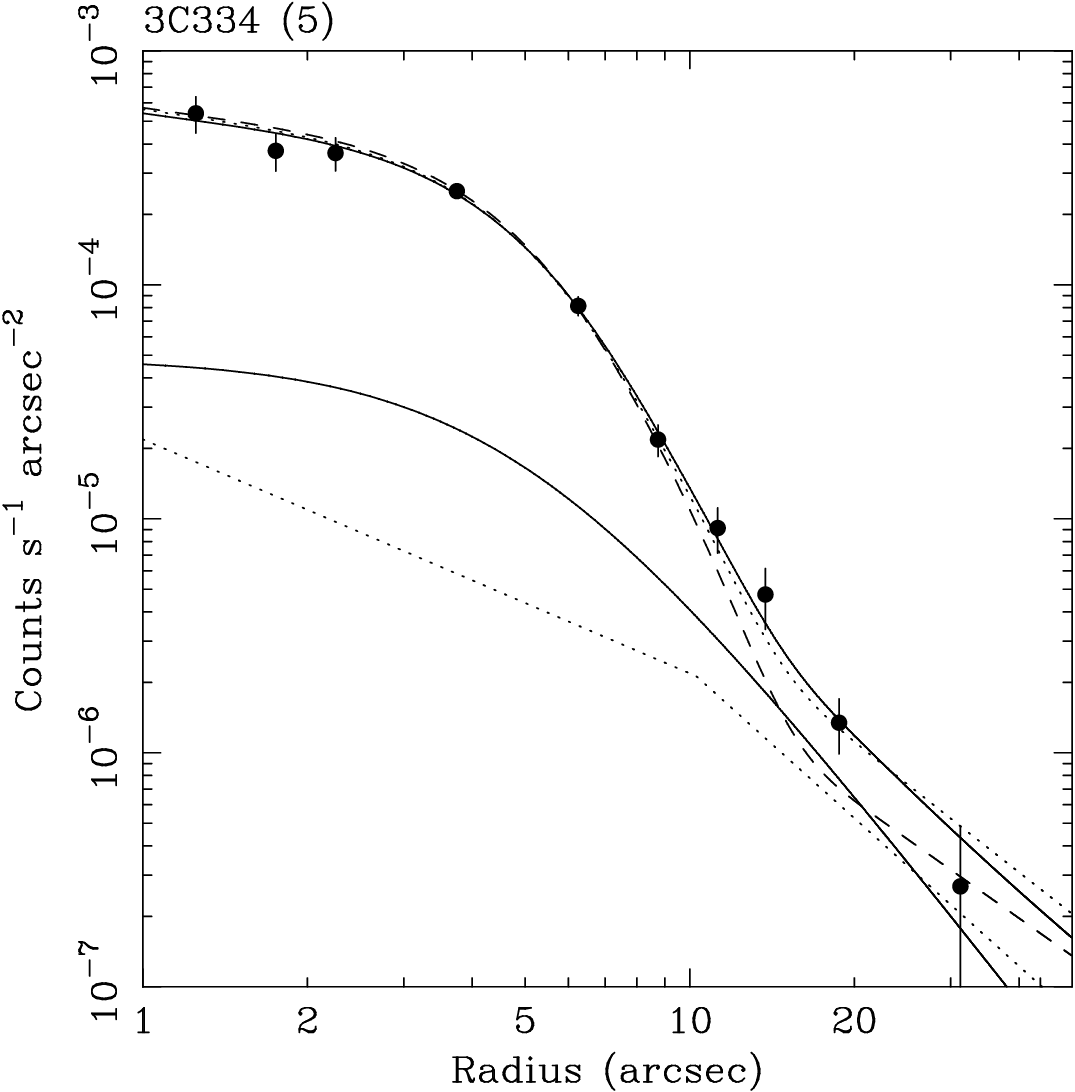,width=0.4\textwidth,angle=270}
\psfig{figure=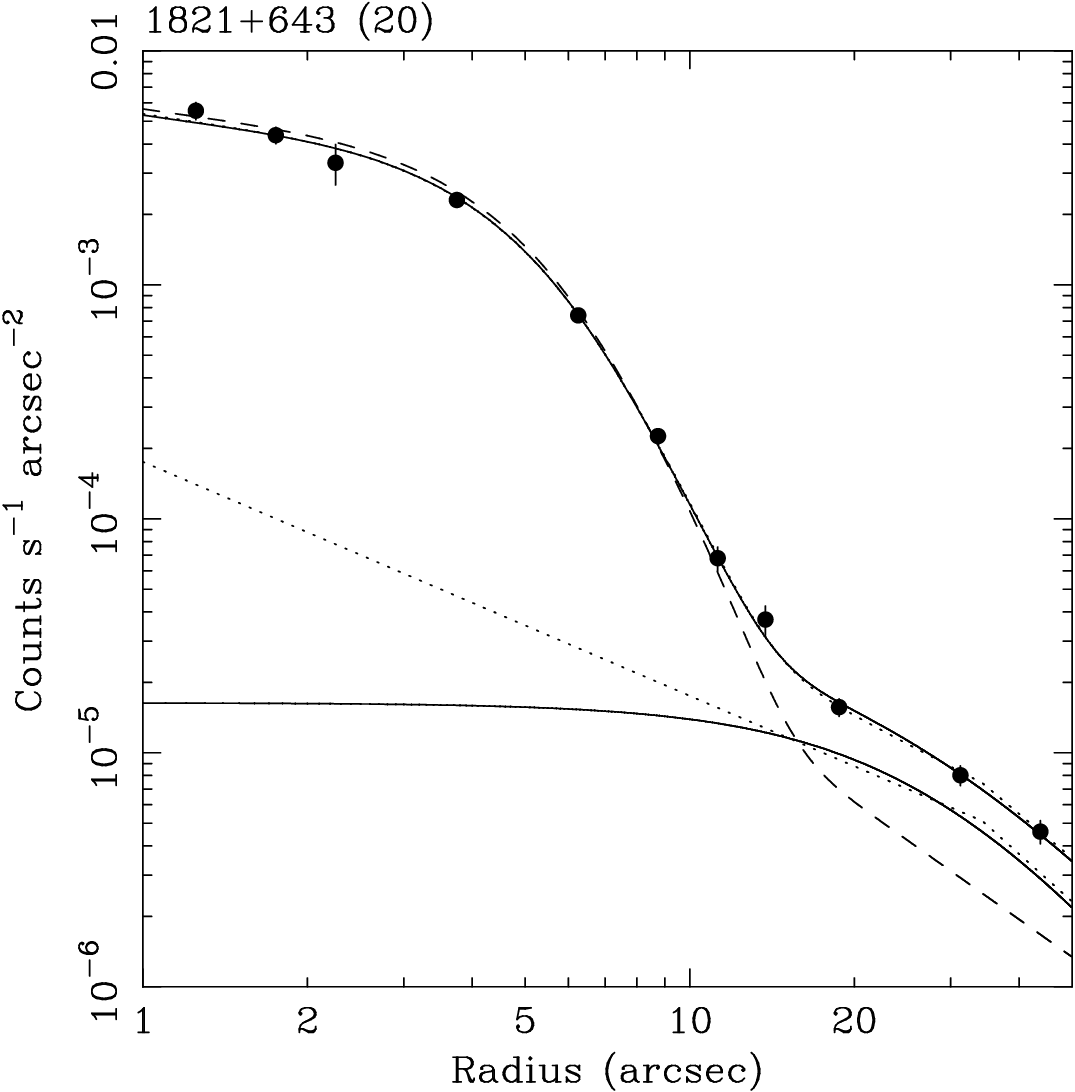,width=0.4\textwidth,angle=270}
\caption{  }
\end{figure}

\subsection {3C275.1} 3C275.1  was the first quasar discovered from optical
galaxy counts to be located at the centre of a rich cluster of
galaxies (Hintzen et al 1981; Hintzen 1984; Ellingson et al 1991). The
radio source is only slightly bent, but the two sides display an
asymmetry in the Faraday depolarization (Garrington et al 1991). The
quasar is embedded in a host galaxy with a continuum spatial profile
and absolute magnitude typical of a bright cluster cD; this is in turn
surrounded by a large (100\kpc) optical emission-line nebula (Hintzen
\& Romanishin 1986; Hintzen \& Stocke 1986). Crawford \& Fabian (1989)
deduce a pressure within this gas of $>3\times10^5$\pcmcuK\ at radii
of $<$20\kpc\ and thus also the presence of an intracluster medium
(Crawford \& Fabian 1989).

  The fit to the X-ray radial profile improves slightly with the
addition of an extended component to the HRI PSF. The models all imply 
a surprisingly broad core, so much so that fitting a second component
to data only within a 50 arcsec radius is not robust. The King and
broken power-law models yield very similar results, with $R\sim$36
arcsec (where 1 arcsec corresponds to 7.4\kpc\ at the redshift of the
quasar), and 24 per cent of the total X-ray luminosity of the source
at $L_X\sim3\times10^{44}$\ergps.

\subsection{3C281}

3C281 is known to lie in a rich cluster (Yee \& Green 1987) and
 Bremer \etal (1992) infer a pressure exceeding 2$\times10^6$\pcmcuK\ 
in the extended emission-line gas within a radius of 20kpc. 
After the wobble-correction has been applied, the X-ray source
associated with the quasar shows a distinct elongation to either side of
the core along a position angle of 45\degmark west of north. The
direction of this elongation is, however, at odds with that of the
radio source which has an axis 10\degmark east of north. 

The profiles extracted from both observations of 3C281 both show an
improved fit from the addition of an extended component (preferably a
King law). The fractional luminosity of this component is high,
ranging over 42-67 per cent at 1.1-1.7$\times10^{45}$\ergps (the
higher values obtained with the King model fits). The characteristic
radius of this component is 5-6 arcsec (where 1 arcsec corresponds to 
7.6\kpc\  at the redshift of the quasar). 

\subsection{3C334}
This quasar lies in a clustered environment (Hintzen 1984) and has a
pressure within the extended emission-line region of over
$6\times10^5$\pcmcuK\ at 30\kpc\ from the quasar core (Crawford
\& Fabian (1989). The quasar again shows a depolarization asymmetry
(Garrington et al 1991), and 
narrow-band imaging by Hes (1995) suggests that the [OII] line
emission is extended along the same position angle of 
$\sim150$\degmark as the strong radio jet to the south-east of the
quasar core. 


The fits to the wobble-corrected profiles show an improvement
to the fit with the addition of an extended component, preferably a
King model. The lengthscale $R$ of this component lies in the range  5-10
arcsec (where 1 arcsec corresponds to 7.4\kpc\  at the redshift of the
quasar), and contains 8-22 per cent of the total X-ray luminosity at
3-8$\times10^{44}$\ergps. 


\subsection{H1821+643 }

H1821+643 is radio-quiet quasar which is luminous in the infrared. The
X-ray emission was found to have a significant extended component by
Hall et al (1997). It is included here as a comparison object. 

The fits improve dramatically with the inclusion of an extended
component. With the 50 arcsec apertures, the King model is often
preferred, whereas the 100 arcsec apertures show a marked improvement
for the broken power-law model. The characteristic radius of this
extended component, $R$, shows a wide range of 29-81 arcsec (where 1
arcsec corresponds to 5.5\kpc\ at the redshift of the quasar), but
clear separation according to both extended model and outer profile
radius employed. The extended component contains from 10--19 per cent
of the X-ray luminosity, at 8--20$\times10^{44}$\ergps.

\section{Summary of results}
We tabulate average properties of the extended emission component from
the profile fits to each quasar in Table~\ref{tab:averages}, where
these averages are derived only from the fits where the normalizations
of the PSF and model for the extended emission are both allowed to
vary (labelled as PSF+broken power-law and PSF+King law in Table~2).
The values are averaged from the fits only to the wobble-corrected
images (and from each phase interval employed). As is clear from both
Tables~2 and \ref{tab:averages}, where the scatter in properties is
sufficiently low for comparison to be made (3C48, 3C215, 3C273 and
3C281), systematic differences can be seen between parameters derived
using the King and broken power-law fits. The King law always gives a
higher X-ray luminosity (and thus higher overall percentage of the
total luminosity), but a smaller characteristic radius $R$ than the
broken power-law model.

The average bolometric luminosities (from the profile fit out to 100
arcsec) span the range of 6-43$\times10^{44}$\ergps, with values of 13
and 18$\times10^{44}$\ergps typical for the broken power-law and King
models respectively. The King models give values of $R$ in the range
33-38\kpc, with the exception of 14\kpc\ for the fits to 3C273; for
the broken power-law model the values are 31-49\kpc. The three
quasars, 3C334, 3C254 and 3C275.1, for which we do not differentiate
between parameters derived from each model have an average $R$ of 49,
94 and 270 \kpc, respectively.

Where a preference between the two models for the extended emission can be
seen, it is nearly always for the King law over the broken power-law model (in 
3C48, 3C273, 3C281 and 3C334). An exception is the 100 arcsec profile fits
to 1821+643, which prefers the broken power-law fits. 
The continuing increase in luminosity of the extended X-ray emission between
the 50 and 100 arcsec profile fits also shows that this component is truly
extended over cluster-wide scales of $\sim400-800$\kpc\ radius.

We note that the discrepancy between the observed, corrected profile
and the PSF is least for 3C273 and 3C48, which are the nearest 3CR
quasars in our sample. If there are systematic errors associated with
the wobble correction that we are unaware of, then these objects will
be the most seriously affected. It is important that the extended
emission from these objects be confirmed with {\em Chandra}.

\addtocounter{table}{+1}
\onecolumn
\begin{table}
\caption{Average properties of the extended component
\label{tab:averages}}
\begin{tabular}{lcccccc}
       &    &  &                    &    &  & \\
Quasar & R (kpc) & \%  & $L_{bol}$ ($10^{43}$\ergps) & R (kpc) & \%  & $L_{bol}$ ($10^{43}$\ergps) \\
outer radius (arcsec) & (50)    & (50) & (50)                   & (100)   & (100) & (100)\\
       &    &  &                    &    &  & \\
3C48 (King) & 33$\pm$4 & 13$\pm$1 & 154$\pm$13 & 35$\pm$2 & 15$\pm$2 & 180$\pm$20 \\
\ \ $''$ \ \ \     (BPL)  & 39$\pm$1 & 11$\pm$1 & 119$\pm$7 & 39$\pm$1 & 11$\pm$1 & 128$\pm$9 \\
3C215 (King) & 32$\pm$1 & 26$\pm$3 & 85$\pm$11 & 33$\pm$1 & 26$\pm$4 & 89$\pm$13 \\
\ \ $''$ \ \ \ \ \    (BPL)  & 49$\pm$3 & 17$\pm$3 & 57$\pm$9 & 49$\pm$3 & 19$\pm$3 & 68$\pm$11 \\
3C254  & 86$\pm$10 & 16$\pm$2 & 121$\pm$9 & 94$\pm$11 & 17$\pm$1 & 141$\pm$4 \\
3C273 (King) & 15$\pm$1 & 10$\pm$1 & 353$\pm$22 & 14$\pm$1 & 12$\pm$1 & 426$\pm$17 \\
\ \ $''$ \ \ \ \ \       (BPL)  & 31$\pm$2 & 5$\pm$1 & 173$\pm$2 & 31$\pm$2 & 5$\pm$1 & 166$\pm2$\\
3C275.1  & (341) & (17)  & (40) & 270$\pm$7  & 24$\pm$1 & 62$\pm$4 \\
3C281 (King) & 39$\pm$1 & 62$\pm$6 & 256$\pm$3 & 38$\pm$1 & 64$\pm$4 & 276$\pm$3 \\
\ \ $''$ \ \ \ \ \       (BPL)  & 50$\pm$2 & 45$\pm$3 & 189$\pm$8 & 48$\pm$5 & 47$\pm$4 & 213$\pm$2 \\
3C334   & 50$\pm$10 & 16$\pm$3 & 106$\pm$18 & 49$\pm$9 & 17$\pm$3 & 117$\pm$18 \\
H1821+643 (King) & 175$\pm$12 & 11$\pm$1 & 218$\pm$11 & 267$\pm$18 & 18$\pm$1 & 380$\pm$9 \\
\ \ $''$ \ \ \ \ \ \ \ \ \ \ \  (BPL)  & 196$\pm$6 & 13$\pm$1 & 254$\pm$11 & 389$\pm$56 & 20$\pm$1 & 431$\pm$11 \\

\end{tabular}
\\
Notes:\\
These values are  averaged only from the results to the fits to
the wobble-corrected images (and each phase
interval employed), and from the PSF+broken power-law and PSF+King law
models. Where the two models give consistently different answers, we
deduce an average for each (eg 3C48); where there is a similar range or 
too few fits available to reliably discriminate for such differences
 (eg 3C275.1) we obtain an average value from both models
together. \\ 
The errors given are the 
variation in this average value and not a significance of the detection. \\
\end{table}

\begin{table}
\caption{Derived cooling flow parameters 
\label{tab:cf}}
\begin{tabular}{lccccc}
 &    &  &  & &  \\
Quasar & $L_{\rm CF}$   & \Mdot & $r_{\rm CF}$ & $P$(30\kpc) & $n(R)$ \\
       & ($10^{43}$\ergps) & (\Msunpyr) & (\kpc) & ($10^6$\yr) & ($10^{-2}$\pcmcu) \\
3C48  & 36 & 330 & 155 & 3.5 & 3.4 \\
3C215 & 18 & 160 & 120 & 2.2 & 1.7 \\
3C254 & 45 & 410 & 165 & 2.7 & 1.2 \\
3C273 & 59 & 510 & 175 & 5.1 & 6.2 \\
3C275.1& 30 & 100 & 110 & 1.4 & 0.2 \\
3C281 & 58 & 510 & 180 & 4.0 & 3.0 \\
3C334  & 33 & 290 & 150 & 3.1 & 2.3 \\
\end{tabular}
\\
Notes:\\
$L_{CF}$ is the X-ray luminosity of the extended component within
the break radius $R$, and  is assumed to be due to a cooling flow in the
cluster. \\
\Mdot is the derived mass cooling rate within radius $R$. \\
$t_{\Lambda}(R)$ is the cooling time at  radius $R$.\\
$P$ is the gas pressure at a radius of 30\kpc\ from the centre of the
cluster.\\
$n(R)$ is the electron density at radius $R$. \\
\end{table}
\twocolumn 

\section{Discussion}

\begin{figure}
\psfig{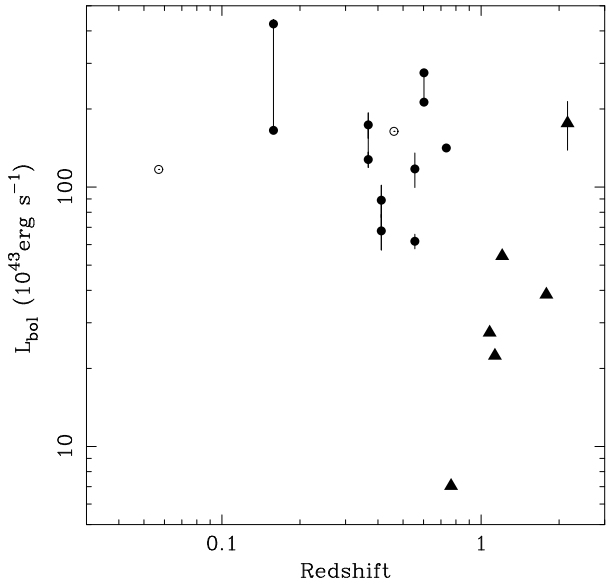}
\caption{ \label{fig:lxz}
The average bolometric  luminosities (as given in
Table~\ref{tab:averages}) for the extended component of emission  (solid
circle markers; values extracted from the 100 arcsec profile fits) plotted
against the redshift of the quasar. Separate averages obtained
for the same quasar assuming
the King or broken power-law models are plotted as two points at the
same redshift joined by a straight line (where the King law gives the upper
end of the range). The luminosities of the host
clusters of the nearby FR~II radio galaxies Cygnus~A and 3C295 are 
plotted as open circles. 
The 
luminosities of the X-ray source associated with the distant radio
galaxies 3C277.2, 3C294,
3C324, 3C356, 3C368 and 1138-262 are also plotted (triangle markers), where
the observed countrates have been converted to X-ray luminosities assuming the
same 4\keV\ thermal bremsstrahlung model used for the quasar extended
emission.  }
\end{figure}

We plot the average bolometric luminosity of the extended cluster
component (from Table~\ref{tab:averages}) against quasar redshift in
Figure~\ref{fig:lxz}. For comparison we plot the bolometric luminosity
of the X-ray source associated with the distant radio galaxies
3C277.2, 3C294, 3C324, 3C356, 3C368 (from Crawford \& Fabian 1996b)
and 1138-262 (Carilli et al 1998), and the clusters surrounding the
two nearby FR~II radio galaxies Cygnus~A (Ueno et al 1994) and 3C295
(Henry \& Henriksen 1986). The observed countrates of the distant
radio galaxies have been converted to luminosities assuming the same
4\keV\ thermal bremsstrahlung model used to obtain luminosities for
the quasar extended emission.

The luminosities we have derived for the environment of our quasars
are brighter than the upper limits of 1.6 -- 3.5$\times10^{44}$\ergps
(rest-frame 0.1-2.4\keV) to any cluster emission surrounding three
radio-loud quasars in Hall et al (1995, 1997). We note, however, that
those upper limits have been obtained from images {\em not} corrected
for satellite wobble. They also assume therefore that the quasar light
follows the standard HRI PSF derived by David et al (1995) and
accounts for all the light in the innermost bin. Our quasar host
clusters are consistent with the luminosity of
3.7$\times10^{45}$\ergps detected by Hall et al (1997) for the
environment of the radio-intermediate quasar H1821+643 at a redshift
$z$=0.297.

The inferred bolometric luminosities of the extended components we
have found here are completely reasonable for moderately rich clusters
of galaxies at low redshift. They are comparable to the luminosities
of the clusters associated with the powerful radio galaxies Cygnus A
and 3C295. They are however (Fig.~1) more luminous than the extended
X-ray emission detected around more distant 3CR radio galaxies above
redshift one. Whether this indicates evolution, a problem for radio
galaxy/quasar unification, or is a result of small number statistics
must await the compilation of a complete sample, which are study is
not.

All extended models have a central cooling time considerably shorter
than a Hubble time. We therefore explore the properties of the implied
cooling flows occurring around these quasars by deriving some
approximate parameters from the broken power-law fits to the profiles.
We attribute all the X-ray luminosity of the extended component within
radius $R$ to thermal bremsstrahlung from gas with electron density
$n$ (where $n\propto r^{-1}$) at a temperature of 4\keV. The cooling
time of the gas at $R$ (except in the case of 3C275.1) is then between
about 1--3 billion yr. We then estimate the cooling flow radius
$r_{\rm CF}$ at which the cooling time is $10^{10}$~yr and obtain a
rough indication of the mass deposition rate within that radius from
the ratio of the mass of gas within $r_{\rm CF}$ to $10^{10}$~yr. The
derived values are shown in Table~\ref{tab:cf}. Note they are of
course subject to not only the appropriateness of the fixed slopes
chosen for our original broken power-law model, but also to the true
gravitational potential of any cluster, and the amount of
gravitational work done on the cooling gas. The values should be
regarded as uncertain by a least a factor of 2. They may be
underestimated by a factor of at least 2 if the gas temperatures are
significantly higher than the 4~keV assumed and there is internal
absorption such as is common in low redshift cooling flows.

Note that the radius of the surface brightness break which we infer is
in the range 40--90~kpc, and is similar to the break radius in the
profile of the cluster around IRAS~09104 ($\sim60$~kpc, Crawford \&
Fabian 1995b). This is likely to be the radius of the core of the
gravitational potential of the cluster; the $r^{-1}$ profile then
occurs within there since that gas is cooling at approximately
constant pressure. Such small gravitational core radii are
characteristic of relaxed lensing cluster cores such as are associated
with massive cooling flows (Allen 1998). The large break radii found
for 3C275.1 and H1821+643 do not agree with this picture and require
more detailed images.

\begin{figure}
\psfig{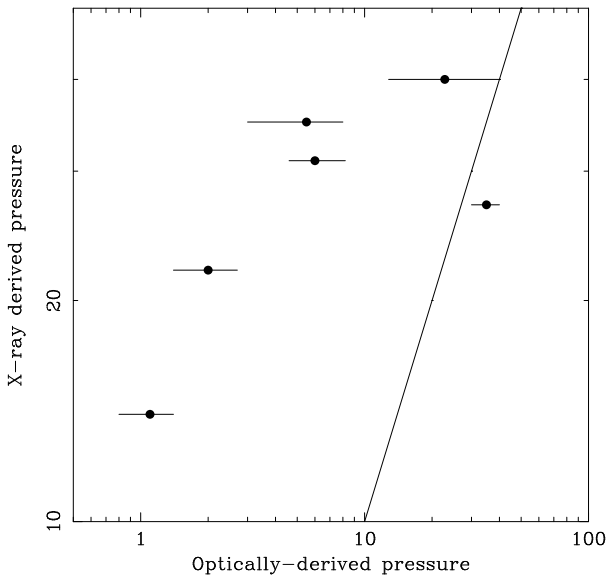}
\caption{ \label{fig:pvsp}
Comparison of the gas pressure at a projected distance of 30\kpc\ from the quasar
derived from the ionization state of the extended emission-line gas, and
from the broken power-law fits to the X-ray radial profiles. Pressures are
expressed in units of $10^5$\pcmcuK, and the solid line indicates the locus
of equal pressures.  }
\end{figure}

We can also use our cooling flow parameters to derive a gas pressure
$P$ at a radius of 30\kpc\ (see Table \ref{tab:cf}) for comparison to
the pressures derived from the completely independent method using the
ionization state of the extended optical emission lines (Crawford \&
Fabian 1989; Bremer et al 1992; Crawford \& Vanderreist 1997). The
pressures derived from the optical nebulosities mostly underestimate
those derived from the X-ray profile fits (Figure~\ref{fig:pvsp}) by a
factor of up to 10. Given that the gas pressures are mostly derived
from the optical nebulosity using a conservative underestimate to the
crucial but unknown UV and soft X-ray band of the ionizing nuclear
spectrum, this discrepancy is not surprising. Support for this
interpretation of the disparity in derived pressures is found in
Crawford et al (1991) where a better knowledge of the ionizing
continuum of the nucleus of 3C263 was found to increase the
optically-derived pressure by up to an order of magnitude. In
addition, we note that 3C254, the quasar with the best agreement
between the two pressure values is the only one where UV HST data has
been used to constrain the shape of the ionizing continuum (Crawford
\& Vanderriest 1997).

We note that the derived cluster luminosity for 3C273 is high and
implies the presence of a rich cluster which is not seen at other
wavelengths. At its relatively low redshift of $z=0.16$, such a
cluster should be obvious in the optical band. As mentioned already,
it has the profile most susceptible to systematic errors, and the
absence of an optical cluster may argue for the existence of such
errors. If the PSF is then uncertain by a relative amount equal to the
observed 3C273 profile and our empirical PSF, then it will not change
greatly our results on the other quasars except perhaps for 3C48.

Inverse-Compton scattering of quasar radiation could still contribute
to our extended component of X-ray emission. If a significant process,
the X-ray source would appear asymmetric and lop-sided, as (back) scattering
by the electrons in the more distant radio lobe should be stronger
than in the nearer lobe. The quality of our current data is
insufficient to show 
any significant asymmetry, and so we are unable to assess the
contribution from this process. Observations of these sources with
{\em Chandra}  and {\em XMM} should clarify its relative importance to
the total emission.

\section{Conclusions}

The seven powerful radio-loud quasars studied here appear to be
surrounded by luminous extended X-ray emission. The spatial properties
of the emission are consistent with an origin in thermal emission from
an intracluster medium. The radiative cooling time of the gas within
$\sim 50$~kpc of the quasars is only a few billion years or less,
indicating the presence of strong cooling flows of hundreds of
$\Msunpyr$. The high pressure of that gas is sufficient to support the
extended optical nebulosities seen around many of these quasars and
may play a r\^ole in shaping the properties of the radio sources, such
as structure, depolarization and possibly even fuelling and evolution
(Fabian \& Crawford 1990). Such a r\^ole is not clear from a comparison of
radio properties with the neighbouring optical galaxy density (Rector
et al 1995).

A small number of luminous clusters without central AGN have so far
been found beyond redshift 0.5, and the discovery of a cluster at
$z\sim0.8$ with luminosity $10^{45}$\ergps (Donahue et al 1998)
provides strong evidence for a low density universe. We have shown
that the study of the environment of powerful radio-loud quasars is a
promising way of extending the discovery of similarly luminous
clusters both in numbers and to higher redshifts.

\section{Acknowledgements}
We thank Steve Allen for advice on the profiles of massive cooling
flows. CSC and ACF thank the Royal Society for financial support. This
work has been supported in part by the DLR (format DARA GmbH) under
grant 50~OR~9403~5 (GH and IL). This research has made use of the
NASA/IPAC Extragalactic Database (NED) and the Leicester Database and
Archive Service (LEDAS).

{}

\begin{figure*}
\vspace{4.25cm}
\hspace{0cm}
\psfig{figure=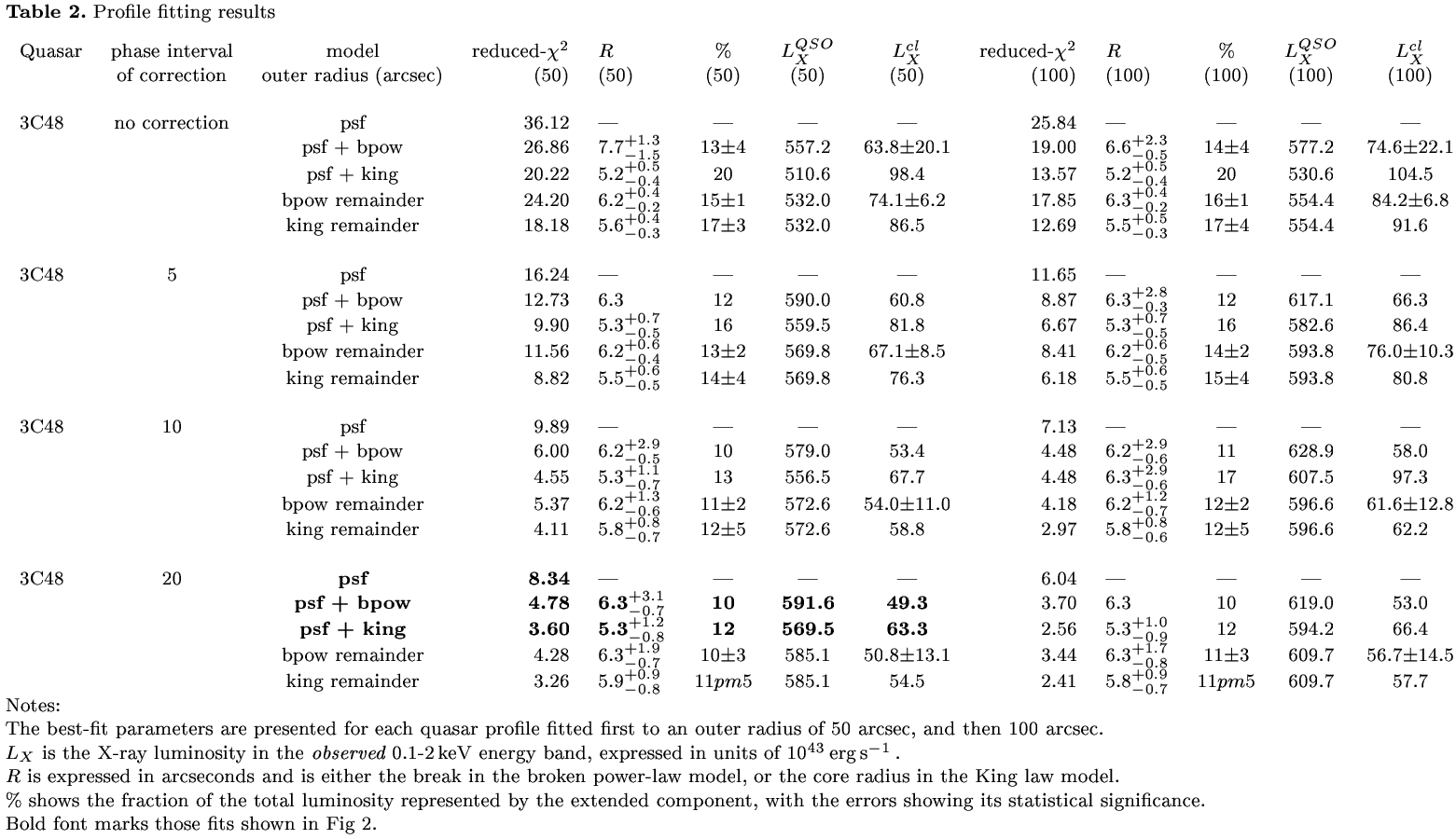,angle=90}
\end{figure*}
\begin{figure*}
\vspace{4.25cm}
\hspace{0cm}
\psfig{figure=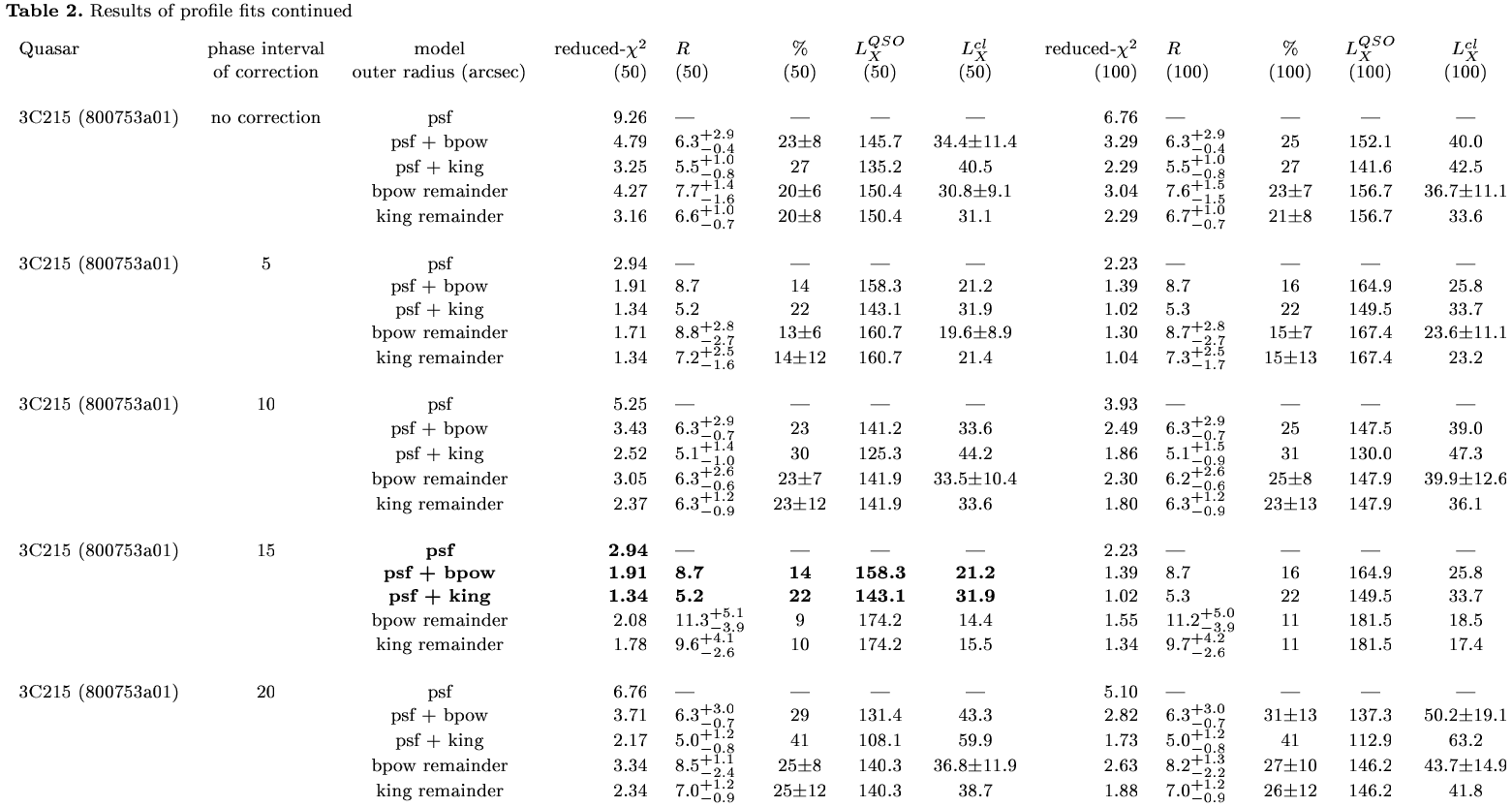,angle=90}
\end{figure*}
\begin{figure*}
\vspace{4.25cm}
\hspace{0cm}
\psfig{figure=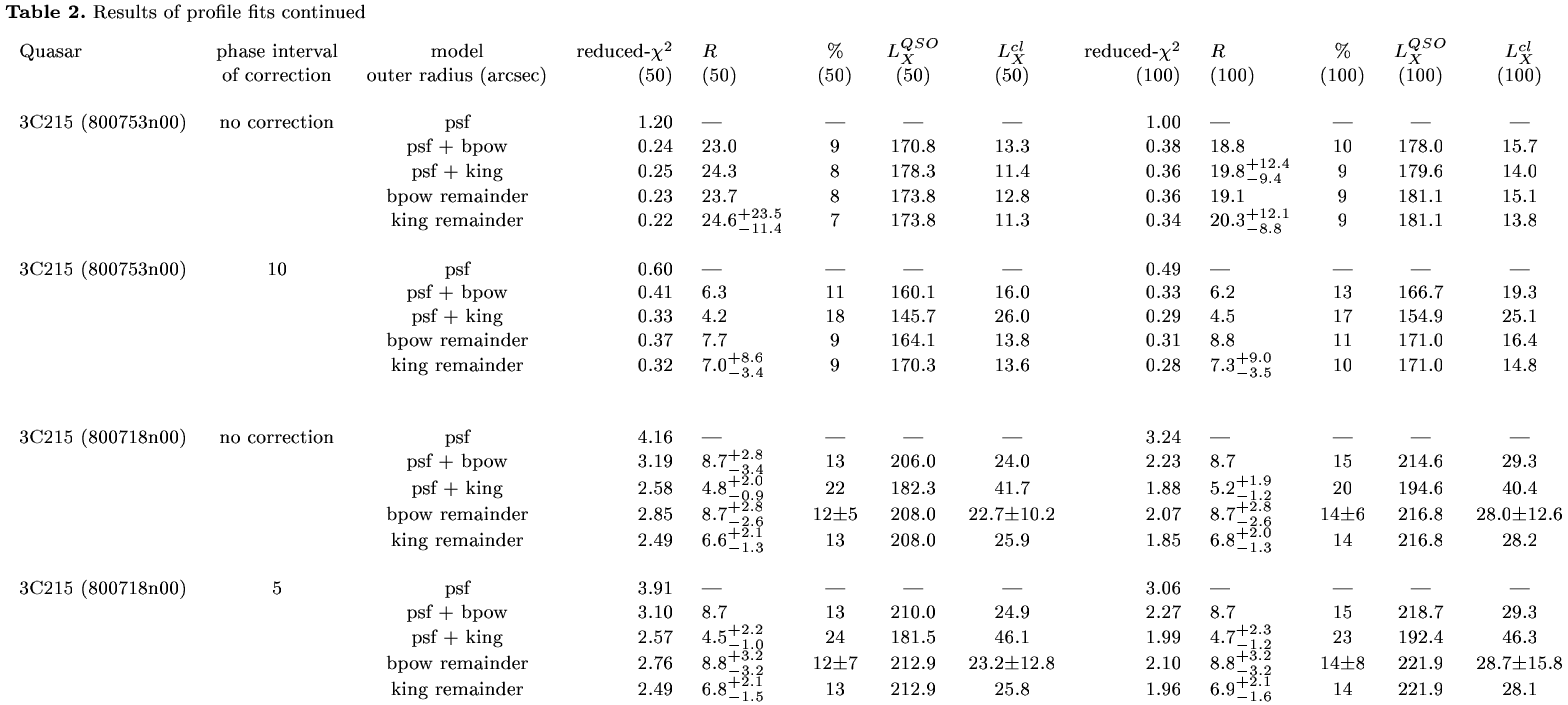,angle=90}
\end{figure*}
\begin{figure*}
\vspace{4.25cm}
\hspace{0cm}
\psfig{figure=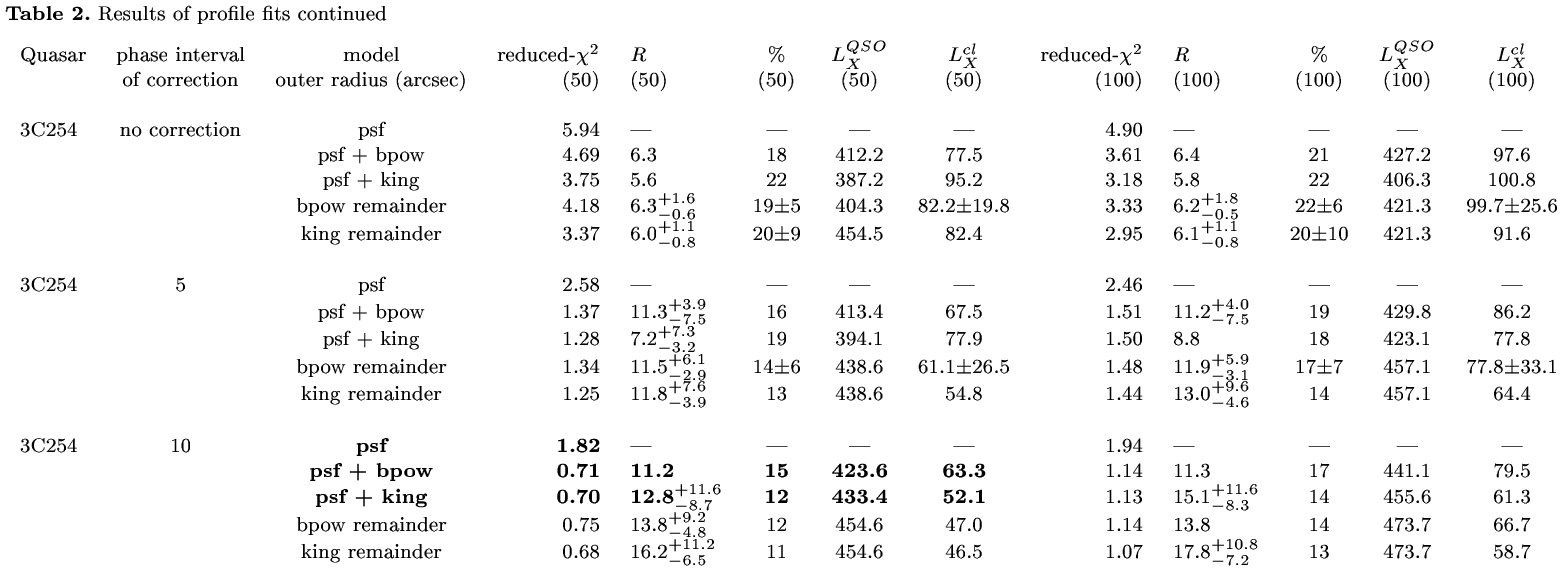,angle=90}
\end{figure*}
\begin{figure*}
\vspace{4.25cm}
\hspace{0cm}
\psfig{figure=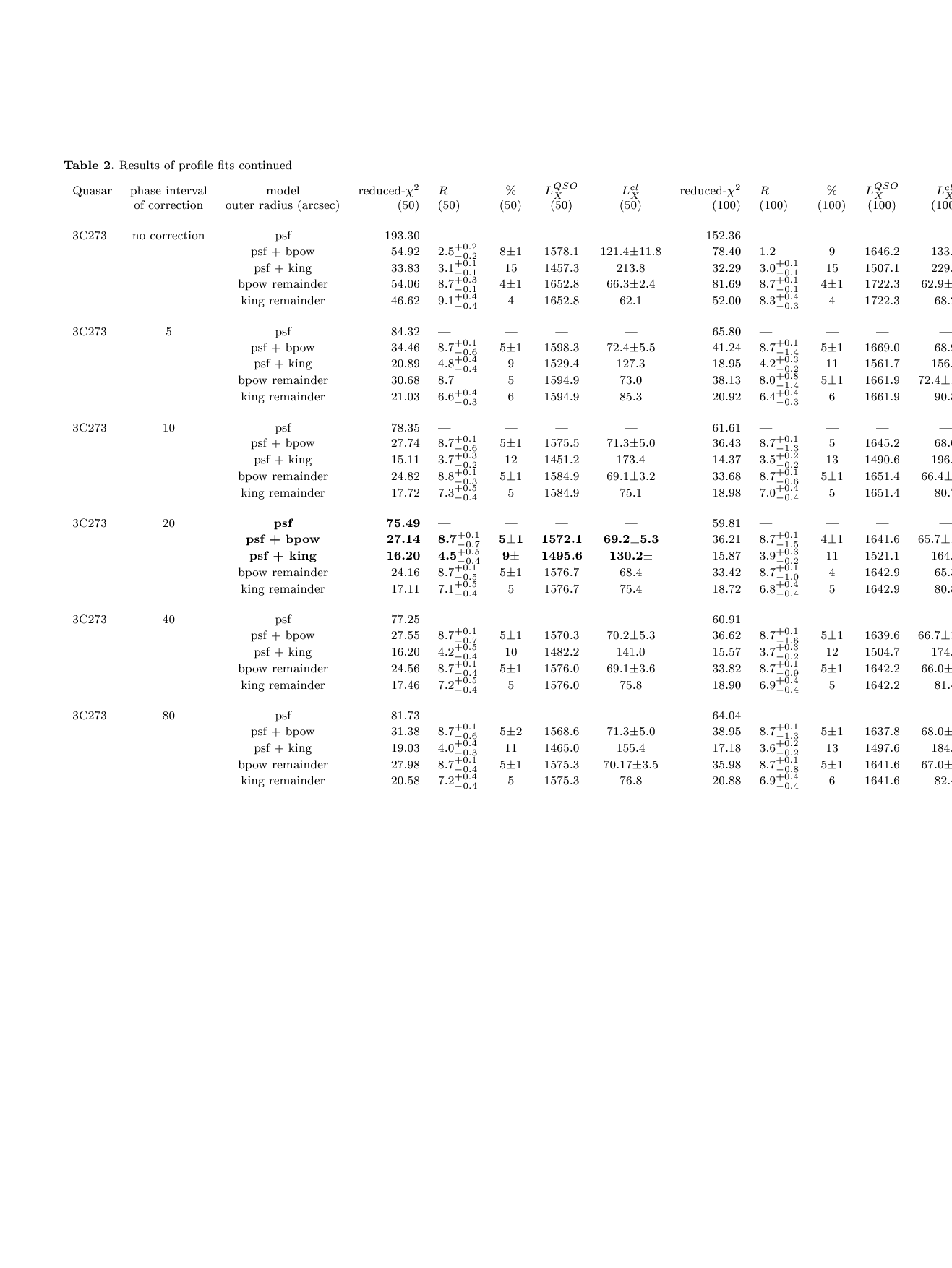,angle=90}
\end{figure*}
\begin{figure*}
\vspace{4.25cm}
\hspace{0cm}
\psfig{figure=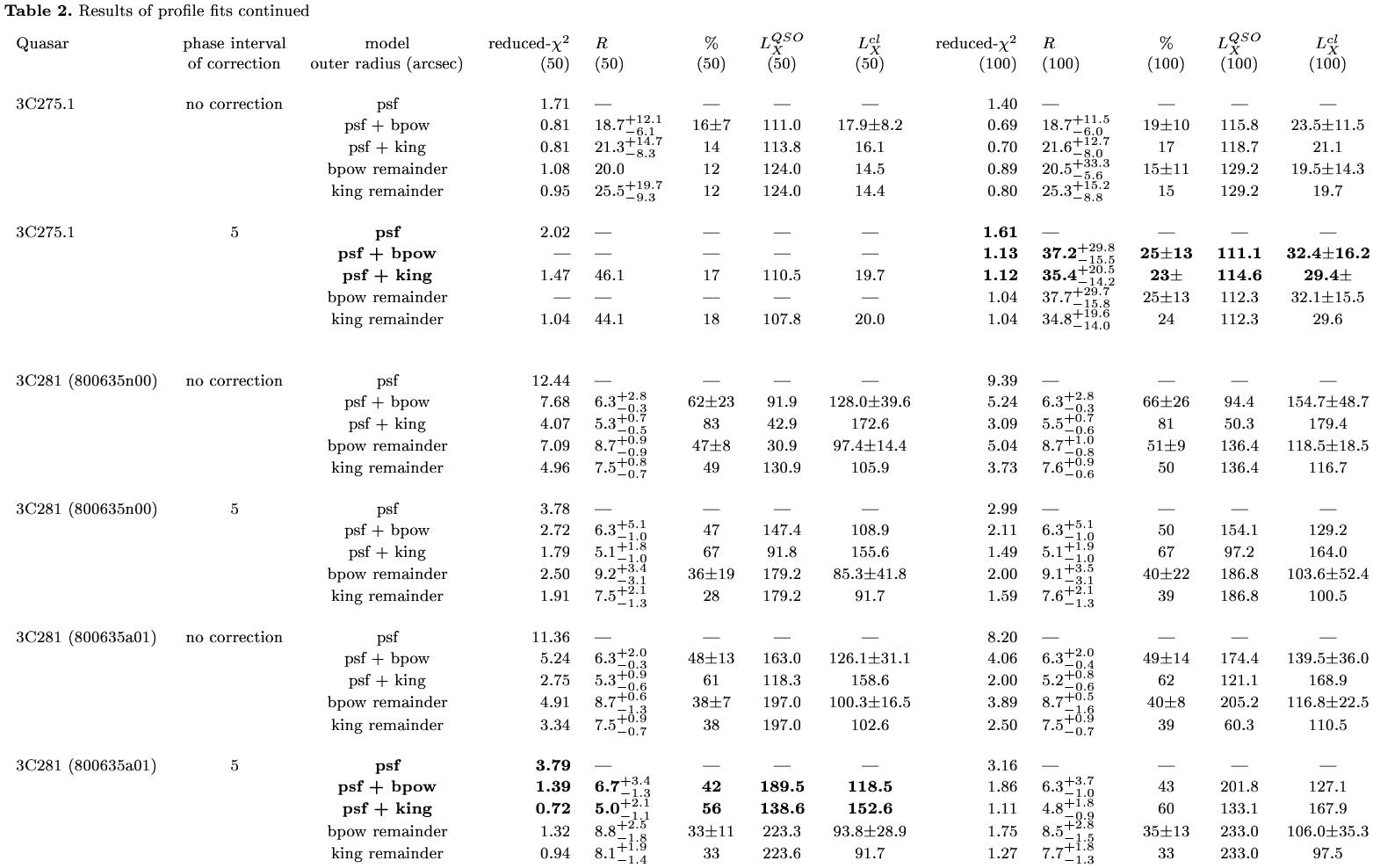,angle=90}
\end{figure*}
\begin{figure*}
\vspace{4.25cm}
\hspace{0cm}
\psfig{figure=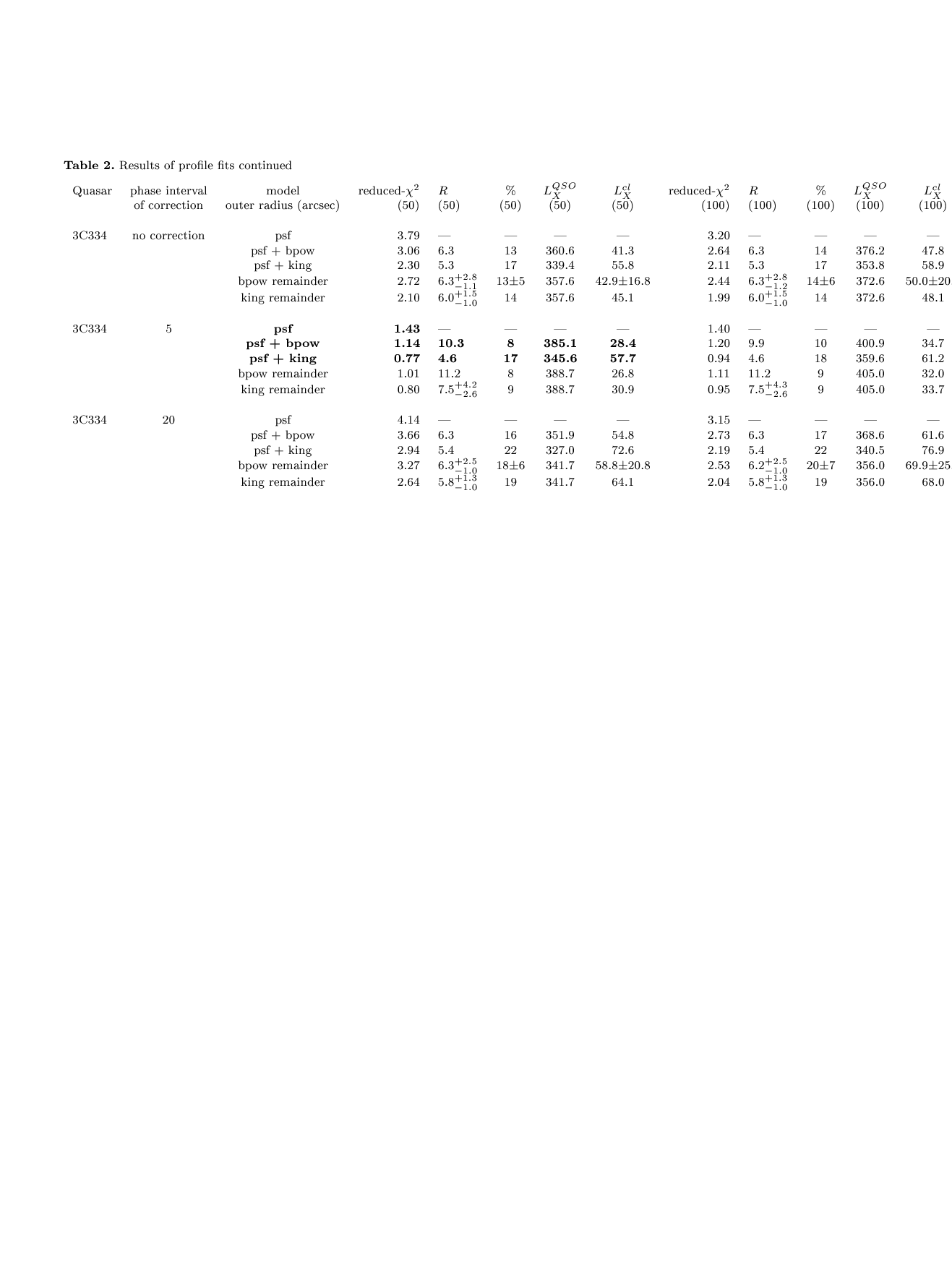,angle=90}
\end{figure*}
\begin{figure*}
\vspace{4.25cm}
\hspace{0cm}
\psfig{figure=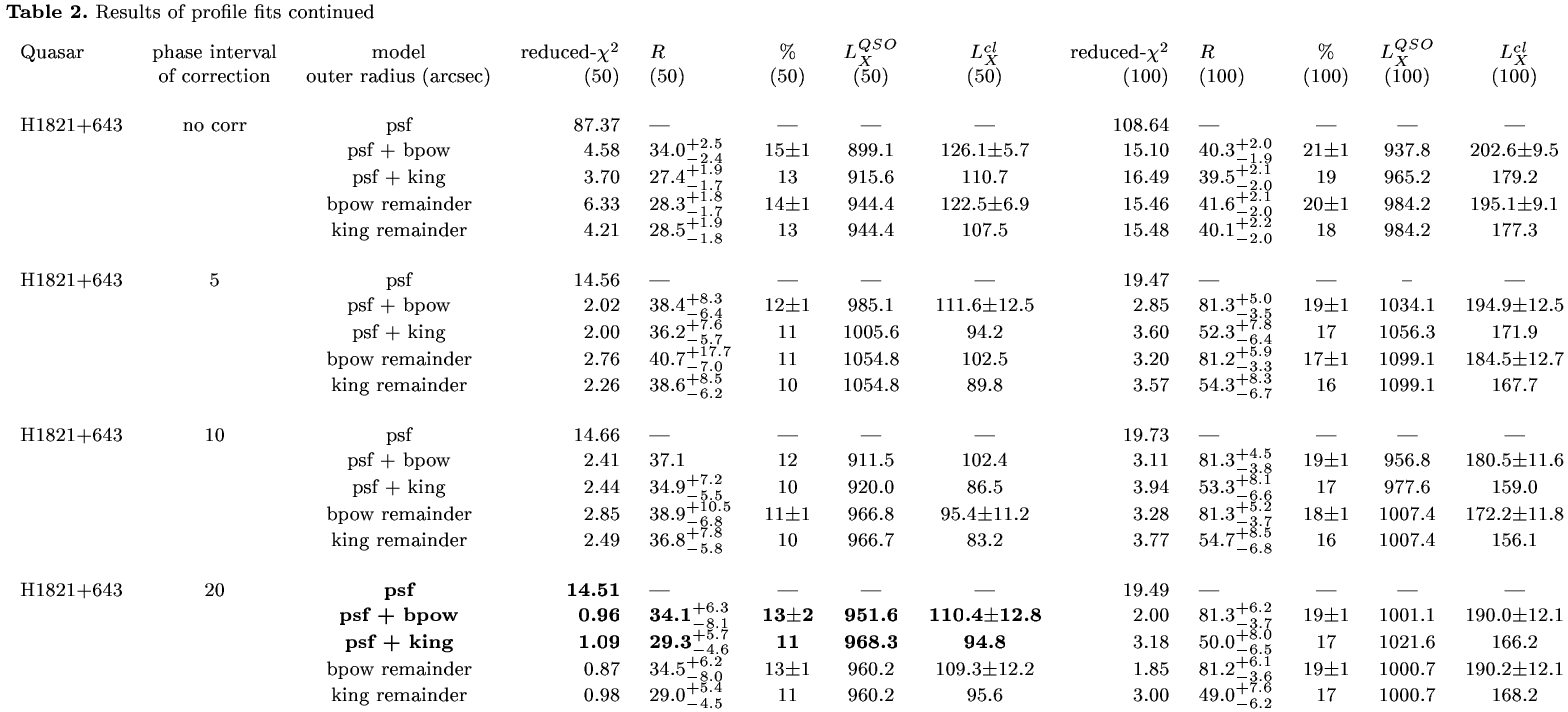,angle=90}
\end{figure*}


\begin{thebibliography}{}

\bibitem [] {} Allen S.W., Fabian A.C., 1997, MNRAS, 286, 583
\bibitem [] {} Allen S.W., 1998, MNRAS, 296, 392
\bibitem [] {} Barthel P.D., Miley G.K., 1988, Nature, 333, 319
\bibitem [] {} Best P, Longair M.S., Rottgering H.J.A., 1998, MNRAS, 295, 549
\bibitem [] {} Bower R, Smail I, 1997, MNRAS, 290, 292
\bibitem [] {} Bremer  M.N., 1997, MNRAS, 284, 126
\bibitem [] {} Bremer  M.N., Crawford C.S., Fabian A.C., Johnstone R.M.,
               1992, MNRAS, 254, 614
\bibitem [] {} Brunetti G., Setti G, Comastri A., 1997, AaA, 235, 898
\bibitem [] {} Carilli C.L., Owen F.N., Harris D.E., 1994, AJ, 107, 480
\bibitem [] {} Carilli C.L. et al, 1997, ApJS, 109, 1
\bibitem [] {} Carilli C.L., Harris D.E., Pentericci L., Rottergering
J.J.A., Miley G.K., Bremer M.N., 1998, ApJ, 494, L143
\bibitem [] {} Crawford CS, 1997, in {\it Observational Cosmology with the
New Radio Surveys}, p99,  eds Bremer M, Jackson N, P\'erez-Fournon, Kluwer, Dordrecht. 
\bibitem [] {} Crawford C.S., Fabian A.C., 1989, MNRAS, 239, 219
\bibitem [] {} Crawford C.S., Fabian A.C., 1993, MNRAS, 260, L15 
\bibitem [] {} Crawford C.S., Fabian A.C., 1995a, MNRAS, 273, 827 
\bibitem [] {} Crawford C.S., Fabian A.C., 1995b, MNRAS, 274, L63 
\bibitem [] {} Crawford C.S., Fabian A.C., 1996a, MNRAS, 281, L5 
\bibitem [] {} Crawford C.S., Fabian A.C., 1996b, MNRAS, 282, 1483 
\bibitem [] {} Crawford C.S., Vanderriest C., 1997, MNRAS, 285, 580
\bibitem []{} David LP, Harnden FR, Kearns KE, Zombeck MV, 1995, {\em The
ROSAT High Resolution Imager}, US ROSAT Science Data Center, Smithsonian
Astrophysical Observatory
\bibitem [] {} Deltorn J.-M., Le Fevre O., Crampton D., Dickinson M.,
1997, ApJ, 483, L21
\bibitem [] {} Dickinson M.,  1997 {\sl HST and the High Redshift
Universe}, eds, Tanvir N., Aragon-Salamanca A.,  Wall J.V., published by World Scientific.
\bibitem [] {} Donahue M., Voit G.M., Gioia I., Lupino G., Huges J.P.,
      Stocke J.T., 1998, ApJ, 502, 550
\bibitem [] {} Durret  F., Pecontal E., Petitjean P., Bergeron J., 1994,
                A\&A, 291, 392
\bibitem [] {} Ebeling H., Edge A.C., Fabian A.C., Allen S.W., Crawford
C.S., B\"ohringer H., 1997, ApJ, 479, L101
\bibitem [] {} Ebeling H. et al 1998, MNRAS, 301, 881
\bibitem [] {} Ellingson E., Yee H.K.C., Green R.F., 1991, ApJS, 76, 455
\bibitem [] {} Fabian A.C., Crawford C.S., Johnstone R.M., Thomas P.A.,
1987,  MNRAS, 228, 963
\bibitem [] {} Fabian A.C., Crawford C.S., 1990, MNRAS, 247, 439
\bibitem [] {} Forbes  D.A., Crawford C.S., Fabian A.C., Johnstone R.M.,
               1990,     MNRAS, 244, 680
\bibitem [] {} Garrington S.T., Conway, R.G., 1991. MNRAS, 250, 198
\bibitem [] {} Garrington S.T., Conway, R.G., Leahy J.P., 1991. MNRAS, 250, 171
\bibitem [] {} Hall P.B, Ellingson E., Green R.F., 1997, AJ, 113, 1179
\bibitem [] {} Hall P.B, Ellingson E., Green R.F., Yee H.K.C, 1995, AJ, 110, 513
\bibitem [] {} Hardcastle M.J., Lawrence C.R., Worrall D.M., 1998, ApJ, 504, 743
\bibitem [] {} Hardcastle M.J. \&  Worrall D.M., 1999, submitted to MNRAS
\bibitem [] {} Harris D.E., Stern C.P., 1987,  ApJ, 313, 136
\bibitem [] {} Harris D.E., Silverman J.D., Hasinger G., Lehmann I., 1998, AaA Suppl, 133, 431
\bibitem [] {} Henry J.P., Henriksen M.J., 1986, ApJ, 301, 689
\bibitem [] {} Hes R., 1995, PhD thesis, University of Groningen
\bibitem [] {} Hill G.J., Lilly S.J., 1991, ApJ,  367, 1
\bibitem [] {} Hintzen P., 1984, ApJSuppl, 55, 533
\bibitem [] {} Hintzen P., Boeshaar G., Scott J., 1981, ApJL, 246, L1 
\bibitem [] {} Hintzen P., Romanishin W, 1986, ApJ, 311, L1
\bibitem [] {} Hintzen P., Stocke J.T., 1986, ApJ, 308, 540
\bibitem [] {} Hintzen P., Ulvestad J., Owen F.N., 1983, ApJ, 88, 709

\bibitem [] {} Lehmann I., Hasinger G., Schwope A.D., Boller Th., 1999,
(astro-ph/9810214)
\bibitem[]{} Liu R., Pooley G., 1991, MNRAS, 249, 343
\bibitem [] {} Morse J.A., 1994, PASP 106, 675
\bibitem [] {} Rector T., Stocke J.T., Ellingson E., 1995, AJ, 110, 1492
\bibitem [] {} Rosati P., Della Ceca R., Norman C., Giacconi R., 1998,
ApJ, 492, 21
\bibitem[]{} Stark A.A., Gammie C.F., Wilson R.W., Bally J., Linke R.A.,
              Heiles C., Hurwitz M., 1992, ApJS, 79, 77
\bibitem [] {} Stockton A., 1980, In {\em Objects of high redshift},
IAU Symp, 92, eds Abell G.O. \& Peebles P.J.E., Dordrecht, Reidel, p89
\bibitem [] {} Ueno S., Koyama K., Nishida M., Yamauchi S., Ward M.J.,
1994, ApJ, 431, L1
\bibitem [] {} Worrall D.M., Lawrence C.R., Pearson T.J., Readhead
A.C.S., 1994, ApJ, 420, L17 
\bibitem [] {} Yates M.G., Miller L., Peacock J.A., 1989, MNRAS 240, 129
\bibitem [] {} Yee H.K.C., Green R.F., 1987, ApJ, 319, 28
\bibitem [] {} Yee H.K.C., Green R.F., Stockman H.S., 1986, ApJSuppl,
62, 681 

\end{thebibliography}
\end{document}